\documentstyle[aps,epsf,preprint]{revtex}
\tightenlines
\begin{document}
\draft

\title{Manifestation of the \lowercase{\boldmath{$a^0_0(980)-f_0(980)$}} mixing
in the reaction \lowercase{\boldmath{$\pi^-p\to\eta\pi^0n$}} on a polarized target\\[0.7cm]}
\author{N.N. Achasov \thanks{Email: achasov@math.nsc.ru}\,\, and
\  G.N. Shestakov \thanks{Email: shestako@math.nsc.ru}\\[.8cm]}
\address{Laboratory of Theoretical Physics,
Sobolev Institute for Mathematics,\\ Novosibirsk, 630090, Russia\\[1cm]}
\maketitle
\begin{abstract}

It is argued that the single-spin asymmetry in the reaction
$\pi^-p\to a_0^0(980)n\to(\eta\pi^0)_S\,n$ is extremely sensitive
to the mixing of the $a^0_0(980)$ and $f_0(980)$ resonances. It is
shown that at low momentum transfers (namely, in any one of the
intervals $0\leq-t\leq0.025,...,0.1$ GeV$^2$) the normalized
asymmetry, which can take the values from $-1$ to 1, must undergo
a jump in magnitude close to 1 in the $\eta\pi^0$ invariant mass
region between 0.965 GeV and 1.01 GeV. The strong asymmetry jump
is the straightforward manifestation of the $a^0_0(980)-f_0(980)$
mixing. For observing the jump, any very high quality $\eta\pi^0$
mass resolution is not required. The energy dependence of the
polarization effect is expected to be rather weak. Therefore, it can be
investigated at any high energy, for example, in the range from eight
to 100 GeV.
\end{abstract}

\vspace*{0.7cm} 

\newpage
\section{Introduction}

Study of the nature of light scalar resonances has become a
central problem of non-perturbative QCD. The point is that the
elucidation of their nature is important for understanding both
the confinement physics and the chiral symmetry realization way in
the low-energy region, i.e., the main consequences of QCD in the
hadron world. The nontrivial nature of the well established
lightest scalar resonances is no longer denied practically
anybody. In particular, there exist numerous evidences in favor of
the four-quark ($q^2\bar q^2$) structure of these states, see, for
example, Ref. \cite{A1} and references therein. In a recent paper
\cite{PRL}, we suggested a new method of the investigation of the
$a_0(980)$ and $f_0(980)$ resonances with use of the polarization
phenomena closely related to the $a_0^0(980)-f_0(980)$ mixing
effect that carries important information on the nature of these
puzzling states, in particular, about their coupling to the $K\bar
K$ channels.

The mixing between the $a_0^0(980)$ and $f_0(980)$ resonances was
discovered theoretically as a threshold phenomenon (induced by the
$K^+$ and $K^0$ meson mass difference) in the late 70s \cite{M1}.
The cross section of the process $\pi^+\pi^-\to\eta\pi^0$, which
is forbidden by $G$-parity but appears because of
$a_0^0(980)-f_0(980)$ mixing, was calculated for the first time
\cite{M1}. Moreover, the reactions $\pi^\pm
N\to\eta\pi^0(N,\Delta)$, $\,K
N\to[(\eta\pi^0),(\pi^+\pi^-)][\Lambda,\Sigma,\Sigma(1385)]$ on
unpolarized targets, the $f_1(1285)\to a_0^0(980)\pi^0\to3\pi$
decay, and the $\bar pn\to(\pi^-,\rho^-)f_0(980)\to(\pi^-,\rho^-)
\eta\pi^0$ annihilations at rest, in which the
$a_0^0(980)-f_0(980)$ mixing could be detected, were considered in
detail, and effects of violation of isospin invariance caused by
this mixing were estimated in the $\eta\pi^0$ and $\pi\pi$ mass
spectra and differential cross sections \cite{M1}.

Recently, interest in the $a_0^0(980)-f_0(980)$ mixing was renewed,
and its possible manifestations in various reactions are
intensively discussed in the literature
\cite{M2,M3,M4,M5,M6,M7,M8,M9,M10,M11,M12,M13,M14,M15,M16,M17}.
However, this mixing has not unambiguously been identified yet in
corresponding experiments. For example, in Ref. \cite{M6} it was
suggested that the data on the centrally produced $a_0^0(980)$
resonance in the reaction $pp\to p_s(\eta\pi^0)p_f$, in principle,
can be interpreted in favor of the existence of $a_0^0(980)-f
_0(980)$ mixing. In Ref. \cite{M9} it was noted that within the
experimental errors and the model uncertainty in the $f _0(980)$
production cross section the result obtained in Ref. \cite{M6}
does not contradict to the predictions made in Ref. \cite{M1}.
However, the experimental confirmation of such a scenario
requires measuring the reaction $pp\to p_s(\eta\pi^0)p_f$ at a
much higher energy to exclude a possible effect of the secondary
Regge trajectories, for which the $\eta\pi^0$ production is not
forbidden by $G$-parity.

A qualitatively new proposal concerning a search for the
$a_0^0(980)-f_0(980)$ mixing effect was given in Ref. \cite{PRL}.
We suggested performing the polarized target experiments on the
reaction $\pi^-p\to\eta\pi^0n$ at high energy in which the fact of
the existence of the $a^0_0(980)-f_0(980)$ mixing can be
unambiguously and very easily established through the presence of
a strong jump in the azimuthal (single-spin) asymmetry of the
$S$-wave $\eta\pi^0$ production cross section near the $K\bar K$
thresholds \cite{PRL}.

In this paper, the expected polarization effect in the reaction
$\pi^-p\to a_0^0(980)n\to(\eta\pi^0)_S\,n$ is considered in more
detail and all the quantitative estimates are discussed as
thoroughly as possible (here and in what follows $(\eta\pi^0)_S$
denotes the $\eta\pi^0$ system with the relative orbital angular
momentum $L=0$). It should be noted that the main features of the
predicted interference pattern are to a great extent model
independent. We show that at low momentum transfers the normalized
asymmetry, which can take the values from $-1$ to 1, must undergo
a jump in magnitude close to 1 in the $\eta\pi^0$ invariant mass
region between 0.965 GeV and 1.01 GeV. The strong asymmetry jump
is the exclusive consequence of isospin breaking due to the
$a_0^0(980)-f_0(980)$ mixing. We also emphasize that observing the
asymmetry jump does not require at all any very high quality
$\eta\pi^0$ mass resolution that would be absolutely necessary to
recognize the $a^0_0(980)-f_0(980)$ mixing manifestation in the
$\eta\pi^0$ mass spectrum in unpolarized experiments. Furthermore,
the energy dependence of the polarization effect is expected to be
rather weak. Therefore, it can be investigated at any high energy,
for example, in the range from eight to 100 GeV.

The paper is organized as follows. In Sec. II, the definitions of
the cross section and asymmetry for the reaction $\pi^-p\to(\eta
\pi^0)_S\,n$ on a polarized target are given. Section III is
purely technical. It contains the detail formulae for the
$\pi^-p\to a_0^0(980)n\to(\eta\pi^0)_S\,n$ reaction amplitudes at
high energies. These amplitudes correspond to the $\rho_2$, $b_1$,
and $\pi$ Regge pole exchange mechanisms. The one-pion exchange
mechanism violates the $G$-parity conservation and arises owing to
the $a^0_0(980)-f_0(980)$ mixing. This section demonstrates also
the specific features of the $a^0_0(980)-f_0(980)$ transition
amplitude, which are of crucial importance for polarization
phenomena. In Sec. IV, the data from the recent unpolarized
experiments on the reaction $\pi^-p\to\eta\pi^0n$ are discussed
and the quantitative estimates of the $\rho_2$ and $b_1$ exchange
contributions to the $(\eta \pi^0)_S$ production cross section are
presented. In Secs. V and VI, the results of the calculation of
the polarization effect due to the $a_0 ^0(980)-f_0(980)$ mixing
are given in the $\rho_2$ and $\pi$ exchange model and in the
$\rho_2$, $b_1$, and $\pi$ exchange one, respectively. The
conclusions based on these results are briefly formulated in Sec.
VII.

\section{Cross section and asymmetry}

Owing to parity conservation, the differential cross section of
the reaction $\pi^-p\to(\eta \pi^0)_S\,n$ on a polarized proton
target at fixed incident pion laboratory momentum, $P^{\pi^-}_{lab
}$, has the form \begin{equation} d^3\sigma/dtdmd\psi=[\,d
^2\sigma/dtdm+I(t,m)\,P\cos\psi\,]/2\pi\,,\end{equation} where $t$
is the square of the four-momentum transferred from the incident
$\pi^-$ to the outgoing $\eta \pi^0$ system, $m$ is the
$\eta\pi^0$ invariant mass, $\psi$ is the angle between the normal
to the reaction plain, formed by the momenta of the $\pi^-$ and
$\eta\pi^0$ system, and the transverse (to the $\pi^-$ beam axis)
polarization of the protons, $P$ is a degree of this polarization,
$d^2\sigma/dtdm=|M_{++}|^2+|M_{+-}|^2$ is the unpolarized
differential cross section, $M_{+-}$ and $M_{++}$ are the
$s$-channel helicity amplitudes with and without nucleon helicity
flip, $I(t,m)=2\,\mbox{Im}(M_{++}M^*_{+-})$ describes the
interference contribution responsible for the azimuthal (or
single-spin) asymmetry of the cross section. In terms of the
directly measurable quantities $I(t,m)$ and $d^2\sigma/dtdm$, one
can also define the dimensionless, normalized asymmetry
$A(t,m)=I(t,m)/[d^2\sigma/dt dm]$, $\,-1\leq A(t,m)\leq1$. The
asymmetry pertaining to some interval of $-t$, $\,A(-t_1\leq-t\leq
-t_2,m)$, or to some interval of $m$, $\,A(t,m_1\leq m\leq m_2)$,
is defined by the ratio of the corresponding integrals of $I(t,m)$
and $d^2\sigma /dtdm$ over $t$ or over $m$. Here we shall be
interested in the region of $m\approx1$ GeV. The available data
from unpolarized target experiments on the reaction
$\pi^-p\to\eta\pi^0n$ \cite{M2,E1,E2,E3,E4} show that the
$(\eta\pi^0)_S$ mass spectrum in this region of $m$ is dominated
by the production of the $a_0^0(980)$ resonance, $\pi^-p\to
a_0^0(980)n\to(\eta\pi^0)_S\,n$.

It follows from the $G$-parity conservation  that at high energies
and small $-t$ the amplitudes $M_{+-}$ and $M_{++}$ are defined by
the $t$-channel exchanges with the quantum numbers of the $b_1$
and $\rho_2$ Regge poles, respectively \cite{M3} (hereinafter they
are denoted by $M^{b_1}_{+-}$ and
$M^{\rho_2}_{++}$).\footnote{History of the $\rho_2$ Regge
exchange has been reviewed in Ref. \cite{M3}. Recall that the
lower-lying representative of the $\rho_2$ Regge trajectory has
the quantum numbers $I^G(J^{PC})=1^+(2^{--})$ which belong to the
$^3D_2$ $q\bar q$ family \cite{PDG}.} In addition, there arises
the possibility of the $\pi$ Regge pole exchange in the reaction
$\pi^-p\to(\eta\pi^0)_S\,n$ by virtue of the process $\pi^-p\to
f_0(908)n\to a^0_0(980)n\to(\eta\pi^0)_S\,n$, which is stipulated
by the $a^0_0(980)-f_0(980)$ mixing violating $G$-parity
\cite{M1,M3}.\footnote{The process $\pi^-p\to f_0(908)n\to
a^0_0(980)n\to(\eta\pi^0)_S\,n$ can also occur via the $a_1$
exchange. The estimates fulfilled on the basis of Ref. \cite{P1}
show, however, that, in this case, the amplitude $M^{a_1}_{++}$
can be neglected in comparison with the other ones for small
$-t$.} As is well known, the amplitude of the $\pi$ exchange in
the reaction $\pi^-p\to(\pi\pi)_S\,n$ is large in the low $-t$
region. Moreover, both the modulus and the phase of the
$a^0_0(980)-f_0(980)$ transition amplitude dramatically change as
functions of $m$ near the $K\bar K$ thresholds. As we shall see,
all of these features lead in the reaction $\pi^-p
\to(\eta\pi^0)_S\,n$ to rather impressive consequences, which can
be easily revealed in polarized target experiments, because they
make possible direct measurements of the interference between the
$\rho_2$ and $\pi$ exchange amplitudes.

\section{Amplitudes of the reaction \lowercase{\boldmath{$\pi^-p\to
 a_0^0(980)n\to(\eta\pi^0)_S\,n$}}}

Let us use the Regge pole model and write the $\rho_2$, $b_1$, and
$\pi$ exchange amplitudes for the reaction $\pi^-p\to
a_0^0(980)n\to( \eta\pi^0)_S\,n$ in the following form:
\begin{equation}
M^{\rho_2}_{++}=e^{-i\pi\alpha_{\rho_2}(t)/2}\,e^{\Lambda_{\rho_2}t/2}
\,(s/s_0)^{\alpha_{\rho_2}(0)-1}\,a_{
\rho_2}\,G_{a_0}(m)\,[2m^2\Gamma_{a_0\eta\pi^0}(m)/\pi]^{1/2}\,,
\end{equation}\begin{equation}
M^{b_1}_{+-}=ie^{-i\pi\alpha_{b_1}(t)/2}\,\sqrt{-t}\,e^{\Lambda_{b_1}
t/2}\,(s/s_0)^{\alpha_{b_1}(0)-1}\,a_
{b_1}\,G_{a_0}(m)\,[2m^2\Gamma_{a_0\eta\pi^0}(m)/\pi]^{1/2}\,,
\end{equation}\begin{equation}
M^{\pi}_{+-}=e^{-i\pi\alpha_{\pi}(t)/2}\,\frac{\sqrt{-t}}{t-m^2_\pi}\,
e^{\Lambda_{\pi}(t-m^2_\pi)/2}\,a_\pi\,e^{i\delta_B(m)}\,G_{a_0f_0}(m)\,[2m^2
\Gamma_{a_0\eta\pi^0}(m)/\pi]^{1/2}\,.\end{equation} It should be
immediately emphasized that the $\pi$ exchange amplitude
$M^\pi_{+-}$, which is forbidden by $G$-parity considerations, is
essentially well known theoretically \cite{M1,M3,P1,P2}. In Eqs.
(2)--(4), $\alpha_j(t)=\alpha_j(0)+\alpha'_j\,t$, $\,a_j$, and
$\Lambda_j/2=\Lambda^0_j/2+\alpha'_j\ln(s/s_0)$ are the
trajectory, residue, and slope of the $j$-th Regge pole [as a
guideline one can accept
$\alpha_\pi(t)\approx0.8(t-m^2_\pi)/\mbox{GeV}^2$, $\,\alpha_{b_1}
(t)\approx-0.21+0.8t/\mbox{GeV}^2$, and
$\alpha_{\rho_2}(t)\approx-0.31+0.8t/\mbox{GeV}^2$], $s\approx2m_p
P^{\pi^-}_{lab}$, $s_0=1$\,GeV$^2$,
$G_{a_0}(m)=D_{f_0}(m)/[D_{a_0}(m) D_{f_0}(m)-\Pi^2_{a_0f_0}(m)]$
is the propagator of the mixed $a^0_0(980)$ resonance \cite{M1},
$\,G_{a_0f_0}(m)=\Pi_{a_0f_0}(m)/[D_{a_0}(m)
D_{f_0}(m)-\Pi^2_{a_0f_0}(m)]$, $\,\Pi_{a_0f_0}(m)$ is the
nondiagonal element of the polarization operator describing the
$a^0_0(980)-f_0(980)$ transition amplitude \cite{M1}, $1/D_r(m)$
is the propagator of an unmixed resonance $r$ with a mass $m_r$,
$\,D_r(m)=m^2_r-m^2+\sum
_{ab}[\mbox{Re}\Pi^{ab}_r(m_{f_0})-\Pi^{ab}_r(m)]$,
$\,r=[a_0(980),f_0(980)]$, $\,ab =(\eta\pi^0,K^+K^-,K^0\bar K^0)$
for $r=a_0(980)$, and $\,ab=(\pi^+\pi^-,\pi^0\pi^0,K^+K^-,K^0\bar
K^0)$ for $r=f_0(980)$, $\,\Pi^{ab}_r(m)$ is the diagonal element
of the polarization operator for the resonance $r$ corresponding
to the contribution of the $ab$ intermediate state \cite{P1,P2},
for $m\geq m_a+m_b$
$$\Pi^{ab}_r(m)=\frac{g^2_{rab}}{16\pi}\left[\frac{m_+m_-}{\pi
m^2}\ln\frac{m_b}{m_a}+\rho_{ab}(m)\left(i-\frac{1}{\pi}\ln
\frac{\sqrt{m^2-m^2_-}+\sqrt{m^2-m^2_+}}
{\sqrt{m^2-m^2_-}-\sqrt{m^2-m^2_+}}\right)\right]\,,$$ where
$g_{rab}$ is the coupling constant of $r$ to the $ab$ channel
(here for identical $\pi^0$ mesons
$g^2_{f_0\pi^0\pi^0}=g^2_{f_0\pi^+\pi^-}/2$),
$\rho_{ab}(m)=[(m^2-m^2_+)(m^2-m^2_-)] ^{1/2}/m^2$, $m_\pm=m_a\pm
m_b$, $m_a\geq m_b$, and
$\Gamma_{rab}(m)=\mbox{Im}[\Pi^{ab}_r(m)]/m
=g^2_{rab}\,\rho_{ab}(m)/16\pi m$ is the width of the $r\to ab$
decay; if $m_-\leq m\leq m_+$, then $(m^2-m^2_+)^{1/2}$ should be
replaced by $i(m^2_+-m^2)^{1/2}$. In Eq. (4), $a_\pi=g_{\pi
NN}\,g_{f_0\pi^+\pi^-}/\sqrt{8\pi}s$, $\,g^2_{\pi
NN}/4\pi\approx14.3$, and $\,\delta_B(m)$ is a smooth and large
phase (of about 90$^\circ$ for $m\approx1$ GeV) of the elastic
background accompanying the $f_0(980)$ resonance in the $S$-wave
reaction $\pi\pi\to\pi\pi$ in the channel with isospin $I=0$
\cite{M1,P1,P2}.

The amplitude of the $a^0_0(980)-f_0(980)$ transition,
$\Pi_{a_0f_0}(m)$, must be determined to a considerable extend by
the $K^+K^-$ and $K^0\bar K^0$ intermediate states \cite{M1}
because of the proximity of the $a ^0_0(980)$ and $f_0(980)$
resonances to the $K\bar K$ thresholds and their strong coupling
to the  $K\bar K$ channels. The sum of the one-loop diagrams
$f_0(980)\to K^+K^-\to a^0_0(9 80)$ and $f_0(980)\to K^0\bar
K^0\to a^0_0(980)$, with isotopic symmetry for coupling constants,
gives \cite{M1}
\begin{eqnarray}
\Pi_{a_0f_0}(m)=\frac{g_{a_0K^+K^-}g_{f_0K^+K^-
}}{16\pi}\Biggl[\,i\,\Bigl(\rho_{K^+K^-}(m)-\rho_{K^0\bar
K^0}(m)\Bigr)\\ \nonumber
\left.-\,\frac{\rho_{K^+K^-}(m)}{\pi}\ln\frac{1+\rho_{K^+K^-}(m)}
{1-\rho_{K^+K^-}(m)}+\frac{\rho_{K^0 \bar
K^0}(m)}{\pi}\ln\frac{1+\rho_{K^0 \bar K^0}(m)}{1-\rho_{K^0\bar
K^0}(m)}\,\right]\,,\end{eqnarray} where $m\geq2m_{K^0}$; in the
region $0\leq m\leq2m_K$, $\rho_{K\bar K}(m)$ should be replaced
by $i|\rho_{K\bar K}(m)|$. The ``resonancelike"\ behavior of the
modulus and phase of the amplitude $\Pi_{a_0f_0}(m)$, induced by
the $K^+$ and $K^0$ meson mass difference, is clearly illustrated
in Figs. 1(a) and 1(b). Note that in the region between the
$K^+K^-$ and $K^0\bar K^0$ thresholds, which is eight MeV wide,
$|\Pi_{a_0f_0}(m)|\approx|g_{a_0K^+K^-}g_{f_0K^+K^-}/16\pi|
[(m_{K^0}^2-m_{K^+}^2)/m_{K^0}^2]^{1/2}\approx0.1265
|g_{a_0K^+K^-}g_{f_0K^+K^-}/16\pi|$, i.e., is of the order of
$m_K\sqrt{m^2_{K^0}-m^2_{K^+}}\approx\sqrt\alpha\,m^2_K$
\cite{M1}.\footnote{It is the unique effect of the $\sqrt{m_d-m_u}\sim
\sqrt\alpha$ order which dominates in our consideration. As for
effects of the $m_d-m_u\sim\alpha$ order, they are small. Such
effects were considered partly in Ref. \cite{M7}, $a_0^0(980)\to
\eta\pi^0\to\pi^0\pi^0\to f_0(980)$. A clear idea of the magnitude
of effects of the $m_d-m_u\sim\alpha$ order gives $|\Pi_{a_0f_0}(m
)|$ at $m<0.95$ GeV and $m>1.05$ GeV, see Fig. 1(a).} From Eqs.
(4) and (5) it follows also that the contribution of $M^\pi _{+-}$
to $\,d^2\sigma/dtdm$, in this mass region, is controlled mainly
by the production of the ratios of coupling constants, i.e.,
$|M^\pi_{+-}|^2\propto\sigma(\pi^+\pi^-\to\eta\pi^0)\propto
(g^2_{f_0K^+K^-}/g^2_{f_0\pi^+\pi^-})(g^2_{a_0K^+K^-}
/g^2_{a_0\eta\pi^0})$.

When constructing the curve for $|\Pi_{a_0f_0}(m)|$ in Fig. 1(a)
and obtaining the quantitative estimates presented below for the
polarization effect, we used the following tentative values of the
$f_0(980)$ and $a_0(980)$ resonance parameters:
$m_{f_0}\approx0.980$ GeV,
$\,g^2_{f_0\pi^+\pi^-}/16\pi\approx\frac{2}{3} \,0.1$\,GeV$^2$,
$\,g^2_{f_0K^+K^-}/16\pi\approx\frac{1}{2}\,0.4$\,GeV$^2$,
$\,\delta_B(m)\approx35.5^\circ+47^\circ m/\mbox{GeV}$,
$\,m_{a_0}\approx0.9847$ GeV, $\,g^2_{a_0K^+K^-}/16\pi\approx
g^2_{f_0K^+K^-}/16\pi\approx\frac{1}{2}\,0.4$\,GeV$^2$, and
$\,g^2_{a_0\eta\pi^0}/16\pi\approx0.25$\,GeV$^2$; in addition, see
also Refs. \cite{M1,PDG,P1,P2,P3,P4,P5,P6}. Figures 1(c) and 1(d)
show the $\pi^+\pi^-$ and $\eta\pi^0$ mass spectra
$dN(f_0(980)\to\pi^+\pi^-)/dm=2m^2\Gamma_{f_0\pi^+
\pi^-}(m)/\pi|D_{f_0}(m)|^2$ and $dN(a_0(980)\to\eta\pi^0)/dm=2m^2
\Gamma_{a_0\eta\pi^0}(m)/\pi|D_{a_0}(m)|^2$ for these values of
the parameters.\footnote{There are many reactions in which a
simplest shape of the solitary $f_0(980)$ resonance line, shown in
Fig. 1(c), is drastically distorted at the expense of the
interference of the $f_0(980)$ with accompanying background
contributions. Thus, the $I=0$ $S$ wave amplitude of the reaction
$\pi\pi\to\pi\pi$ in the $f_0(980)$ region has the form
$T^0_0=(e^{2i\delta_B(m)}-1)/2i
+e^{2i\delta_B(m)}m\Gamma_{f_0\pi\pi}(m)/D_{f_0}(m)$, where the
phase of the smooth, elastic background, $\delta_B(m)$, is close
to 90$^\circ$. It is this circumstance that leads to the fact that
the $f_0(980)$ resonance is observed in the corresponding
$\pi\pi\to\pi\pi$ cross section as an interference dip. In our
case, the amplitude $M^\pi_{+-}$ includes the amplitude of the
reaction $\pi^+\pi^-\to f_0(980)\to K\bar K\to
a_0^0(980)\to\eta\pi^0$, which along with the resonance phase must
possess the additional phase of the elastic nonresonant
background in the $\pi\pi$ channel. That is the reason why the
factor $e^{i\delta_B(m)}$ was introduced in Eq. (4). In similar
situations, this is the simplest (and the conventional) way to
account for the nonresonance contributions in accordance with the
unitarity condition. Since the phase $\delta_B(m)$ is large, it is
very important to take it into account. We know nothing about an
analogous background phase in the $\eta\pi^0$ channel. However, in
this case, the phase is common for the amplitudes in Eqs. (2)--(4),
and hance it is absolutely inessential. The
additional details to the aforesaid discussion see, for example,
in Refs. \cite{M1,P1,P2,P3}.} Note that, while $\Gamma_{f_0\pi\pi}
(m_{f_0})=\frac{3}{2}\,\Gamma_{f_0\pi^+\pi^-}(m_{f_0})
\approx98$ MeV and $\Gamma_{a_0\eta\pi^0}(m_{f_0})\approx166$ MeV,
the visible (effective) widths of the corresponding peaks at their
half maxima are approximately equal to $42$ MeV and $68$ MeV,
respectively. This is so owing to the couplings of the resonances
to the $K\bar K$ channels. In its turn, Figs. 1(e) and 1(f) give
an idea of the absolute values and the typical shapes of the
differential cross sections due to the $\pi$ exchange,
$d\sigma^\pi/dt=\int|M^\pi_{+-}|^2dm$ and
$d\sigma^\pi/dm=\int|M^\pi_{+-}|^2dt$, corresponding to the
integration regions over $m $ from 0.8 to 1.2 GeV and over $t$
from $-0.025$\,GeV$^2$ to 0, respectively, and
$P^{\pi^-}_{lab}=18.3$ GeV (i.e., the Brookhaven National
Laboratory (BNL) energy \cite{M2}), at
which $ \Lambda_\pi/2\approx4.5$\,GeV$^{-2}$ \cite{S1,S2}.
Integrating $d\sigma^\pi/dt$ presented in Fig. 1(e) over $t$, we
find that the total cross section of the reaction $\pi^-p\to
a_0^0(980)n\to( \eta\pi^0)_S\,n$ caused by the $\pi$ exchange,
$\sigma^\pi$, is approximately equal to 10.9\,nb. Based on the
previous investigations \cite{M1,M3}, one can conclude that the
indicated value of $\sigma^\pi$ should be considered as its rather
reliable lower bound. Thus, using the above values of the
$a_0^0(980)$ and $f_0(980)$ resonance parameters, we present the
most conservative estimates of the expected polarization effect.
Finally, it should be particularly emphasized that the sharp and
strong variation (by about $90^\circ$) of the phase of the
amplitude $\Pi_{a_0f_0}(m)$ between the $K\bar K $ thresholds,
being crucial for polarization phenomena, is generally independent
of the $f_0(980)$ and $a^0_0(980)$ resonance parameters [see Fig.
1(b) and Eq. (5)].

To estimate quantitatively the $\rho_2$ and $b_1$ exchange
contributions to the $(\eta \pi^0)_S$ production cross section,
and to clarify a question about the relative role of the $\pi$
exchange, it is necessary to turn to the available experimental
data.

\section{Data from unpolarized experiments}

The experiments on the reaction $\pi^-p\to\eta\pi^0n$ on
unpolarized targets were performed at $P^{\pi^-}_{lab}=18.3$ GeV
at BNL \cite{M2,E1,E2}, 38 GeV at Institute High Energy Physics
(IHEP, Protvino) \cite{E3,E4}, 32 GeV at
IHEP \cite{E4}, and 100 GeV at CERN \cite{E4}. The present
situation is rather interesting. The point is that, in general,
the available data from BNL \cite{M2}, IHEP \cite{E3,E4}, and CERN
\cite{E4} do not require at all the introduction of the $b_1$
exchange amplitude $M^{b_1}_{+-}$ to describe the $t$
distributions ($dN/dt$) of the $\pi^-p\to
a_0^0(980)n\to(\eta\pi^0)_S\,n$ reaction events in the
$a_0^0(890)$ mass region. All the data for $0\leq
-t\leq(0.6-0.8)\,$GeV$^2$ are excellently approximated by the
simplest exponential form $C\exp( \Lambda t)$ \cite{M3,E3,E4}
corresponding to the amplitude $M^{\rho_2}_{++}$ nonvanishing for
$t\to0$ \cite{M3}. For example, such a fit to the normalized BNL
data \cite{M2,M3} for the differential cross section $d\sigma/dt$
of the reaction $\pi^-p\to a^0_0(980)n\to( \eta\pi^0)_S\,n$, shown
in Fig. 2 by the solid curve, gives $\chi^2/n.d.f.=15.75/22$ and
$d\sigma/dt=[(945.8\pm46.3)\mbox{nb/GeV}^2]\exp[t(4.729
\pm0.217)/\mbox{GeV}^2]$. Here it is necessary to clarify that the
experimental points shown in Fig. 2 correspond to the BNL data for
$dN/dt$ at $P^{\pi^-}_{lab}=18.3$\,GeV \cite{M2} normalized to the
$a^0_2(1320)$ formation cross section in the reaction $\pi^-p\to
a_2^0(1320)n$ in such a way as it was done in Ref. \cite{M3}.
According this estimate the total cross section $\sigma$ for the
reaction $\pi^-p\to a^0_0(980)n\to(\eta\pi^0 )_Sn$ at 18.3 GeV is
approximately equal to $200$\,nb \cite{M3}. This number we refer
to the $m$ region from 0.8 to 1.2 GeV and to the whole region of
$t\leq0$. Note that the above value of $\sigma$ is in close
agreement with the estimate presented in Ref. \cite{E3}. Comparing
the indicated values of $\sigma$ and $d\sigma/dt$ with the values
of $\sigma^\pi$ and $d\sigma^\pi/dt$ estimated in the previous
section\,\footnote{The curve for $d\sigma^\pi/dt$ shown in Fig.
1(e) has been reproduced also in Fig. 2 for convenience of the
comparison of $d\sigma/dt$ and $d\sigma^\pi/dt$.}, we obtain that
$\sigma^\pi\approx10.9$\,nb makes up about 5.5\% of the total
reaction cross section $\sigma\approx200$\,nb, and that $d
\sigma^\pi/dt$ at the maximum, located near $t\approx-0.0149$
GeV$^2$, $\approx139$ nb/GeV$^2$ accounts for approximately 14.7\%
of $(d\sigma/dt)|_{t\approx0}$, see Fig. 2. However, the main
point is that the whole value of $d\sigma^\pi/dt$ at given $t$, in
fact, comes from the narrow region of $m$ near the $K\bar K$
thresholds, see Figs. 1(a) and 1(f), whereas the values of the
total differential cross section $d\sigma/dt$ are assembled over
the $m$ region which is at least by an order of magnitude wider,
see, for example, Fig. 1(d). Thus, at low $-t$ and $m$ near the
$K\bar K$ thresholds, the $\pi$ exchange contribution can be quite
comparable with that of the $G$-parity conserving $\rho_2$
exchange.

Certainly, the $b_1$ exchange contribution cannot be fully
rejected only on the basis of a good quality of the fit to the
measured $t$ distributions \cite{M2,E3,E4} with the simplest
function $C\exp(\Lambda t)$. In principle, the unpolarized data
\cite{M2,E3,E4} are fitted equally well by using the expression
$dN/dt=C_1\exp(\Lambda_1t)-tC_2\exp(\Lambda_2t)$ \cite{M3,E3},
where the first term can be identified with the contribution of
the amplitude $M^{\rho_2} _{++}$ and the second one with that of
the amplitude $M^{b_1}_{+-}$, see Eqs. (2) and (3). The particular
example of such an approximation of the BNL data is shown in Fig.
2. The dotted curve in this figure presents
$d\sigma/dt=d\sigma^{\rho_2}/dt+d\sigma^{b_1}/dt$, where
$d\sigma^{\rho_2}/dt=958.1(\mbox{nb/GeV}^2)
\exp(7.6t/\mbox{GeV}^2)$ and
$d\sigma^{b_1}/dt=-t\,2486.6(\mbox{nb/GeV}^4)\exp(5.8t/\mbox{GeV}^2)$
are shown by the long and short-dashed curves, respectively. For
this fit $\chi^2=15.7$ and, as is seen, the dotted curve
practically coincides with the solid one corresponding to the fit
using the $\rho_2$ exchange model, for which $\chi^2=15.75$. In
this example, the $b_1$ exchange yields approximately 37\% of the
integrated cross section. Thus, the $\chi^2$ test does not allow
to select between the two models.\footnote{Let us emphasize that
only polarized target experiments would allow to elucidate
unambiguously a question about the contribution of the $b_1$
exchange amplitude because they would make possible direct
measurements the interference term $\mbox{Im}(M_{++}M_{+-}^*)$,
together with the sum $|M_{++}|^2+|M_{+-}|^2$.} The additional
information on the $\rho_2$ and $b_1$ exchange model can be
obtained by fitting to the BNL data with use of the following
parametrization:
\begin{equation} d\sigma/dt=\sigma[(1-B_{b_1})
\Lambda_1e^{\Lambda_1t}-tB_{b_1}\Lambda_2^2e^{\Lambda_2t}]\,,
\end{equation} where $\sigma=200$ nb,
$\Lambda_1$ and $\Lambda_2$ are free parameters, and $B_{b_1}$ is
a portion of the integrated cross section caused by the $b_1$
exchange. The resulting values of $\chi^2$, $\Lambda_1$, and
$\Lambda_2$ as functions of $B_{b_1}$ are plotted in Fig. 3; the
marked points on the $\chi^2$ and $\Lambda_1$ curves correspond
(from left to right) to the values of $B_{b_1}=0$, 0.37, 0.61,
0.7, 0.8, and 0.9. As is seen from Fig. 3(a), $\chi^2$ has a wide
plateau. It remains practically unchanged with increasing
$B_{b_1}$ from 0 to 0.37 \footnote{The $\chi^2$ passes through its
``invisible" minimum at $B_{b_1}\approx0.37$.} and rises by one
only when $B_{b_1}$ reaches 0.61. At $B_{b_1}=0.7$ \,
$\chi^2\approx19.6$ and then increases very rapidly with
$B_{b_1}$, see Fig. 3(a). The approximation of the BNL data
obtained in the fit using Eq. (6) at $B_{b_1}=0.7$ is shown in
Fig. 4. Formally, this approximation is very similar to the
approximations shown in Fig. 2 for $B_{b_1}=0$ (the solid curve)
and $B_{b_1}=0.37$ (the dotted curve). The crucial difference
between these examples is, however, in the values of the slope
$\Lambda_1$ corresponding to the $\rho_2$ exchange contribution.
Figure 3(b) shows the very strong increase of $\Lambda_1$ with
$B_{b_1}$. If the values of $\Lambda_1\approx4.729$ GeV$^{-2}$ (at
$B_{b_1}=0$) and $\Lambda_1\approx7.6$ GeV$^{-2}$ (at
$B_{b_1}=0.37$) are quite reasonable from the Regge phenomenology
standpoint for the secondary Regge exchanges, then the slope
$\Lambda_1\approx17.3$ GeV$^{-2}$ (at $B_{b_1}=0.7$) is already
very unlikely, as well as its larger values. Nevertheless, for
completeness sake we consider the polarization effect in this case
also, see Sec. VI.

\section{Polarization effect in the
\lowercase{\boldmath{$\rho_2$}} and \lowercase{\boldmath{$\pi$}}
exchange model}

Taking into account the above uncertainty of information on the
$b_1$ exchange contribution, it is reasonable to present in the
first place the results of the calculation of the polarization
effect due to the $a_0 ^0(980)-f_0(980)$ mixing in the model
including only the $\rho _2$ and $\pi$ exchange mechanisms.

In Figs. 5(a)--5(c) are shown
$d\sigma/dm=\int[|M^{\rho_2}_{++}|^2+|M^\pi_{+-}|^2]dt$,
$\,d\sigma^{\rho_2}/dm=\int|M^{\rho_2}_{++}|^2dt$, $\,I(m)=\int
I(t,m)dt=\int2\mbox{Im}[M^{\rho_2}_{++}(M^\pi_{+-})^*]dt$,
pertaining to the $-t$ interval from 0 to 0.025 GeV$^2$ at
$P^{\pi^-}_{lab}=18.3$ GeV, and the corresponding asymmetry
$A(0\leq-t\leq0.025\mbox{\,GeV}^2,\,m)$. Figures 5(d)--5(f) show
the same values but pertaining to the the $-t$ interval from 0 to
0.2 GeV$^2$. In so doing, the parameters of the $\rho_2$ exchange,
which we substitute in Eq. (2), correspond to the above-mentioned
fit to the BNL data, shown by the solid curve in Fig. 2. Notice
that the $I(m)$ and asymmetry are determined only up to the sign
because the relative sign of the $\rho_2$ and $\pi$ exchanges is
unknown. Let us also note that, in the $\rho_2$ and $\pi$ exchange
model, the values $I(m)$ and asymmetry for the reaction
$\pi^-p\to(\eta\pi^0)_S\,n$ and for the charge-symmetric reaction
$\pi^+n\to(\eta\pi^0)_S\,p$ are equal in magnitude but opposite in
sign. Figure 5 shows that the polarization effect caused by the
interference between the amplitudes $M_{++}^{\rho_2}$ and
$M_{+-}^\pi$ is quite considerable in any one of the intervals
$0\leq-t\leq0.025,...,0.2$ GeV$^2$. A natural measure of the
effect is the magnitude of a distinctive jump of the asymmetry,
which takes place in the $m$ region from 0.965 to 1.01 GeV. As is
seen from Figs. 5(c) and 5(f), the corresponding difference
between the maximal and minimal values of the asymmetry smoothed
at the expense of the finite $\eta\pi^0$ mass
resolution\,\footnote{ Smoothing the initial values $d\sigma/dt$
and $I(m)$ had been made by using a Gaussian distribution with the
dispersion of ten MeV. The measured mass distributions are always
smoothed by finite resolutions of spectrometers. Therefore, for
instance, the dotted curves in Fig. 5 can be directly compared
with corresponding experimental histograms having approximately
10-MeV-wide mass step and a good statistical accuracy. Obtaining
similar high quality data in the unpolarized $\eta\pi^0$
production experiments has already become commonplace
\cite{E1,E2,E3,E4}.} turns out to be approximately equal to 0.95
in this mass region for the interval $0\leq-t\leq0.025$ GeV$^2$,
see the dotted curve in Fig. 5(c), and $\approx0.75$ for the
interval $0\leq-t\leq0.2$ GeV$^2$, see the dotted curve in Fig.
5(f). It is quite clear that the jump of the asymmetry is the
exclusive consequence of the sharp variation, by 90$^\circ$, of
the phase of the $a^0_0(980)-f_0(980)$ transition amplitude
between the $K^+K^-$ and $K^0\bar K^0$ thresholds. The large
magnitude of the jump is due to both the considerable value of the
modulus of the amplitude $\Pi_{a_0f_0}(m)$ and the enhancement of
its manifestation owing to the $\pi$ exchange mechanism.

Note that any noticeable variation of the interference pattern
does not arise if one refits the BNL data in Fig. 2 by adding the
$\pi$ exchange contribution, indicated in the same figure, to the
$\rho_2$ exchange one.

\section{Polarization effect in the
\lowercase{\boldmath{$\rho_2$}}, \lowercase{\boldmath{$b_1$}}, and
\lowercase{\boldmath{$\pi$}} exchange model}

Let us now see what is changed by including the $b_1$ exchange
contribution.

In the first place we note that if the reaction $\pi^-p\to
a_0^0(980)n\to(\eta\pi^0)_S\,n$ is determined only by the $\rho_2$
and $b_1$ exchange mechanism, then this would lead to a rather
cheerless $m$ dependence of $A(t,m)$. The asymmetry in this case
would be independent of $m$ in the $a_0(980)$ peak region for any
$t$. Indeed, the phase of the production $M_{++}(M_{+-})^*$ in the
$\rho_2$ and $b_1$ exchange model is defined only by the Regge
signature factors of the amplitudes $M^{\rho_2}_{++}$ and
$M^{b_1}_{+-}$, see Eqs. (2) and (3), and
$A(t,m)=\pm\cos[\pi(\alpha_{\rho_2}(t)-\alpha_{b_1}(t))/2]
\times2|M^{\rho_2}_{++}||M^{b_1}_{+-}|/[|M^{\rho_2}_{++}|^2+|M^{
b_1}_{+-}|^2]$, where the factors involving the resonance $m$
dependence simply cancel. Here $\pm$ denotes that the relative
sign of the $\rho_2$ and $b_1$ exchanges is unknown. However, it
is clear that the absolutely different feature should be expected
in the presence of the amplitude $M_{+-}^\pi$: for low $-t$,
$\,A(t,m)$ as a function of $m$ must sharply vary near the $K\bar
K$ thresholds.

We performed the calculations in the $\rho_2$, $b_1$, and $\pi$
exchange model using the example of the fit to the BNL data
described in Sec. IV and shown in Fig. 2. Recall that, in this
example, the $b_1$ exchange contribution makes up approximately
37\% of the total cross section. The polarization effect
corresponding to this case is demonstrated in Fig. 6. The solid
(and dotted) curves in this figure show the asymmetry
$A(0\leq-t\leq-t_2,\,m)$, pertaining to three $-t$ intervals,
$0\leq-t\leq0.025, 0.1, 0.2$\,GeV$^2$, without (and with) a
Gaussian smearing (with the dispersion of ten MeV). The left and
right parts of Fig. 6 correspond to the different choices of the
sign of the $b_1$ exchange amplitude. The overall sign of the
asymmetry is unknown and was chosen arbitrarily as well as in the
case of the $\rho_2$ and $\pi$ exchange model. Note that if the
interference pattern corresponding, for example, to the left
(right) part of Fig. 6 is realized for the reaction
$\pi^-p\to(\eta\pi^0)_S\,n$, then, for the charge-symmetric
reaction $\pi^+n\to(\eta\pi^0)_S\,p$, must be realized that
corresponding to the right (left) part of Fig. 6, but with the
opposite sign. Figure 6 clearly shows that the asymmetry
pertaining to any interval of $0\leq-t\leq0.025,...,0.1$\,GeV$^2$,
as before, undergos a jump of order one in the region $0.965\leq
m\leq1.01$ GeV owing to the $\pi$ exchange admixture. However, the
jump takes place now not relatively to the the zeroth value of the
asymmetry but relatively to its value determined by the
interference between the $\rho_2$ and $b_1$ exchange
contributions. Thus, Figs. 5 and 6 together give already a quite
exhaustive idea of the polarization effect caused by the
$a^0_0(980)-f_0(980)$ mixing, which should be expected in the
reaction $\pi^-p\to a_0^0(980)n\to(\eta\pi^0)_S\,n$. It should be
mentioned in addition to this pattern that the polarization effect
is found to be large in the low $-t$ region, as is clear from Fig.
7, even for such a practically improbable case when the $b_1$
exchange contribution makes up 70\% of the total cross section
(see discussion in Sec. IV).

Let us make yet some general remarks. First, it should be
particularly emphasized that the reliable observation of the
asymmetry jump does not require at all any very high quality
$\eta\pi^0$ mass resolution that would be absolutely necessary to
recognize the $a^0_0(980)-f_0(980)$ mixing manifestation in the
$\eta\pi^0$ mass spectrum in unpolarized experiments. Really, the
fine structure arising in $d\sigma/dm$ by the
$a^0_0(980)-f_0(980)$ mixing is strongly shaded by a mass
smearing, see Figs. 5(a) and 5(d), but in so doing the asymmetry
jump remains, as is clear from Figs. 5(c), 5(f), 6, and 7, in
spite of some smearing. Secondly, by virtue of the theoretically
expected (and experimentally supported!) nearness of the $\pi$,
$\rho_2$, and $b_1$ Regge trajectories, the energy dependence of
the considered polarization effect (the asymmetry magnitude)
should be expected to be rather weak. Therefore, the effect can be
investigated, in fact, at any high energy, for example, in the
range from eight to 100 GeV. Moreover, even if we slightly err,
guiding by the simplest Regge pole model in constructing the
$G$-parity conserving amplitudes (for example, in the choice of
their phases), we are sure of that a jump of a single-spin
asymmetry in the reaction $\pi^-p\to
a_0^0(980)n\to(\eta\pi^0)_S\,n$ in the $\eta\pi^0$ invariant mass
region between the $K^+K^-$ and $K^0\bar K^0$ thresholds will
necessarily take place owing to the specific $m$ dependence of the
$a^0_0(980)-f_0(980)$ transition amplitude [see Figs. 1(a) and
1(b)], its enhancement due to the one-pion exchange mechanism of
the $f_0(980)$ production, and the suppression of the $b_1$
exchange amplitude in the low $-t$ region.

\section{Conclusion}

Thus, we conclude that the interference between the amplitudes
$M_{++}$ and $M_{+-}$ in the reaction $\pi^-p\to
a_0^0(980)n\to(\eta\pi^0)_S\,n$ at small $-t$, which can be
measured in polarized target experiments, turns out to be
extremely sensitive to the mixing of the $a_0^0(980)$ and
$f_0(980)$ states. The asymmetry jump near the $K\bar K$
thresholds is the direct consequence of the $a^0_0(980)-f_0(980$
mixing, and even very rough indications in the presence of such a
jump will allow to draw inferences about the existence of the
mixing effect.

Currently, experimental investigations utilizing the polarized
beams and targets are on the rise. Therefore, this analysis seems
to be quite opportune. The relevant experiments on the reaction
$\pi^-p\to\eta\pi^0n$ on a polarized proton target, in principle,
can be realized at High Energy Accelerator Research Organization
(KEK, Tsukuba), BNL, IHEP, CERN (COMPASS), Fermi National
Accelerator Laboratory (Batavia), Institute of Theoretical and
Experimental Physics (Moscow), and
Institut f\"ur Kernphysik in J\"ulich. Discovery of the
$a^0_0(980)-f_0(980)$ mixing would open one more interesting page
in investigation of the nature of the puzzling $a^0_0(980)$ and
$f_0(980)$ states. Of course, the general idea of using
polarization phenomena as an effective tool for the observation of
the $a_0^0(980)-f_0(980)$ mixing connected with a great variation
(by about 90$^\circ$) of the phase of the $a^0_0(980)-f_0(980)$
mixing amplitude in the narrow energy region (8 MeV) between the
$K^+K^-$ and $K^0\bar K^0$ thresholds is also applicable to other
reactions.

\section{ACKNOWLEDGEMENTS}

This work was supported in part by the RFBR Grant No. 02-02-16061
and the Presidential Grant No. 2339.2003.2 for support of Leading
Scientific Schools.

\newpage
\begin{figure}\centerline{\epsfysize=7.5in\epsfbox{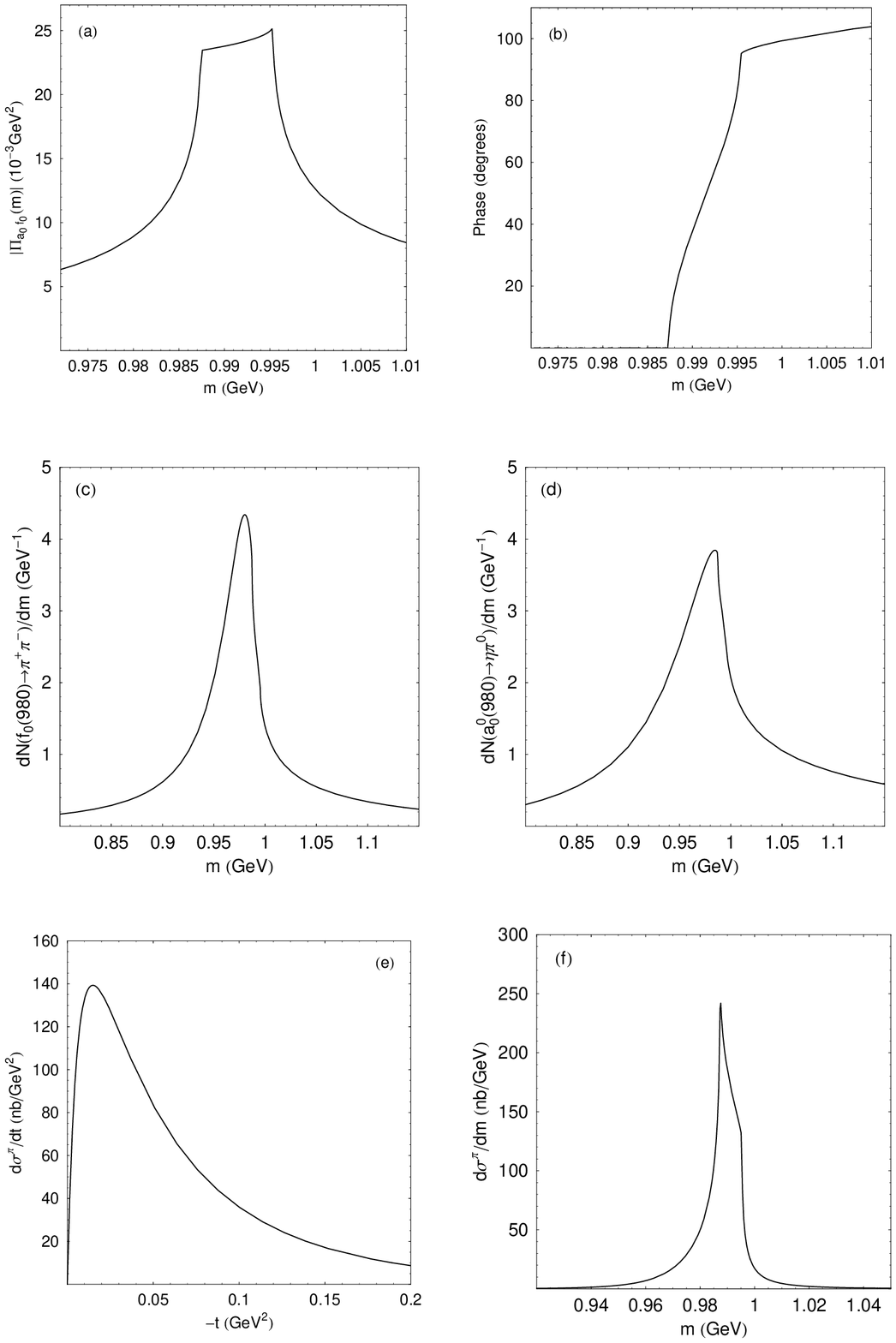}}
\caption{(a) An example of the modulus of the
$a^0_0(980)-f_0(980)$ transition amplitude $\Pi_{a_0f_0}(m)$, see
Eq. (5). (b) The phase of the $a^0_0(980)-f_0(980)$ transition
amplitude $\Pi_{a_0f_0}(m)$, see Eq. (5). (c) An example of the
$\pi^+\pi^-$ mass spectrum corresponding to the solitary
$f_0(980)$ resonance. (d) An example of the $\eta\pi^0$ mass
spectrum corresponding to the solitary $a^0_0(980)$ resonance. (e)
The differential cross section $d\sigma^\pi/dt$ for the reaction
$\pi^-p\to f_0(980)n\to a^0_0(980)n\to(\eta\pi^0)_S\,n$, due to
the $\pi$ exchange mechanism, at $P^{\pi^-}_{lab}= 18.3$ GeV and
for the region $0.8\leq m\leq1.2$ GeV. (f) $d\sigma^\pi/dm$ for
the same reaction and $P^{\pi^-}_{lab}$ corresponding to the
interval $0\leq-t\leq0.025$ GeV$^2$.}\end{figure}

\begin{figure}\centerline{\epsfysize=5in\epsfbox{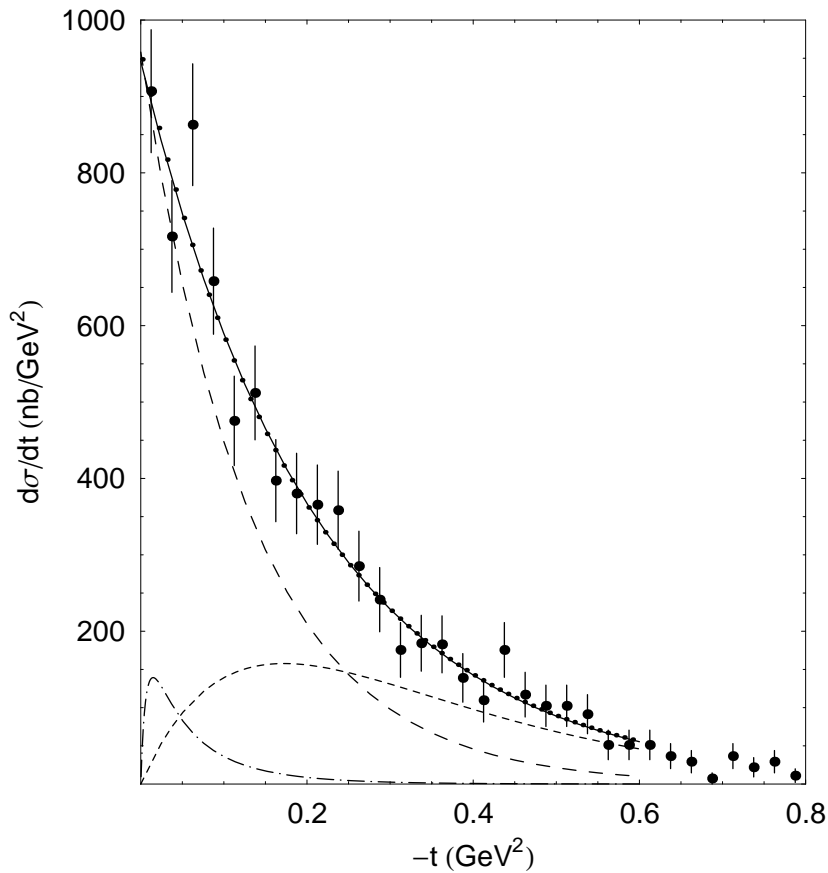}}
\caption{The experimental points are the normalized BNL data [4]
for $d\sigma/dt$ of the reaction $\pi^-p\to a^0_0(980)n\to(
\eta\pi^0)_S\,n$ at $P^{\pi^-}_{lab}= 18.3$ [5]. The solid curve
corresponds to the best fit of the data obtained in the $\rho_2$
exchange model. The dotted curve, which practically coincides with
the solid one, gives an example of the approximation of the data
with use of the $\rho_2$ and $b_1$ exchange model; in so doing,
the long-dashed and short-dashed curves show the $\rho_2$ and
$b_1$ exchange contributions to $d\sigma/dt$, respectively. The
dot-dashed curve is the differential cross section
$d\sigma^\pi/dt$ corresponding to Fig. 1(e).}\end{figure}

\begin{figure}\centerline{\epsfysize=4.5in\epsfbox{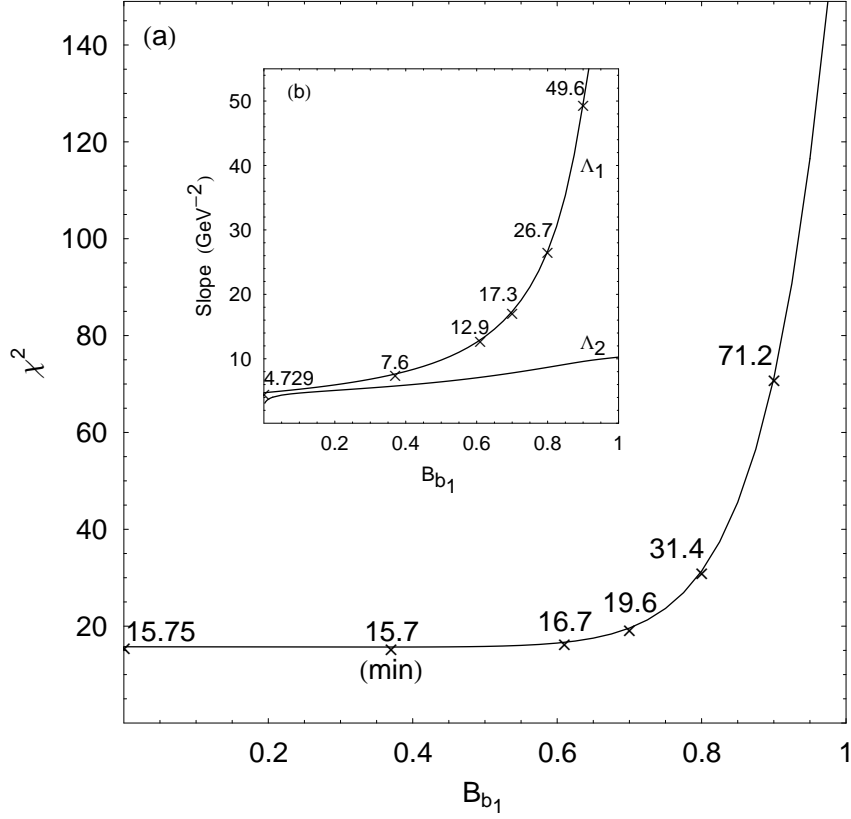}}
\caption{The $\chi^2$, $\Lambda_1$, and $\Lambda_2$ as functions
of $B_{b_1}$ obtained by fitting to the BNL data (shown in Fig. 2)
with use of Eq. (6). The marked values of $\chi^2$ in (a) and
$\Lambda_1$ in (b) correspond (from left to right) to the values
of $B_{b_1}=0$, 0.37, 0.61, 0.7, 0.8, and 0.9.}\end{figure}

\begin{figure}\centerline{\epsfysize=5in\epsfbox{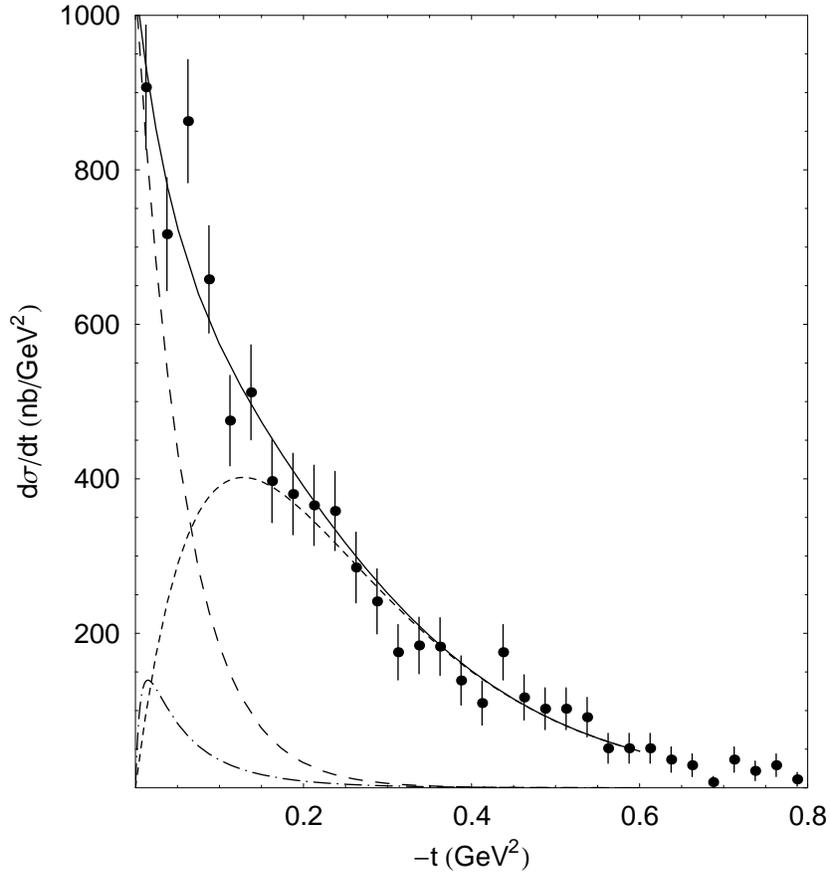}}
\caption{The experimental data and the dot-dashed curve are the
same as in Fig. 2. The solid curve corresponds to the fit to the
data using Eq. (6) at $B_{b_1}=0.7$. The long-dashed and
short-dashed curves show the $\rho_2$ and $b_1$ exchange
contributions to $d\sigma/dt$, respectively.}\end{figure}

\begin{figure}\centerline{\epsfysize=7.5in\epsfbox{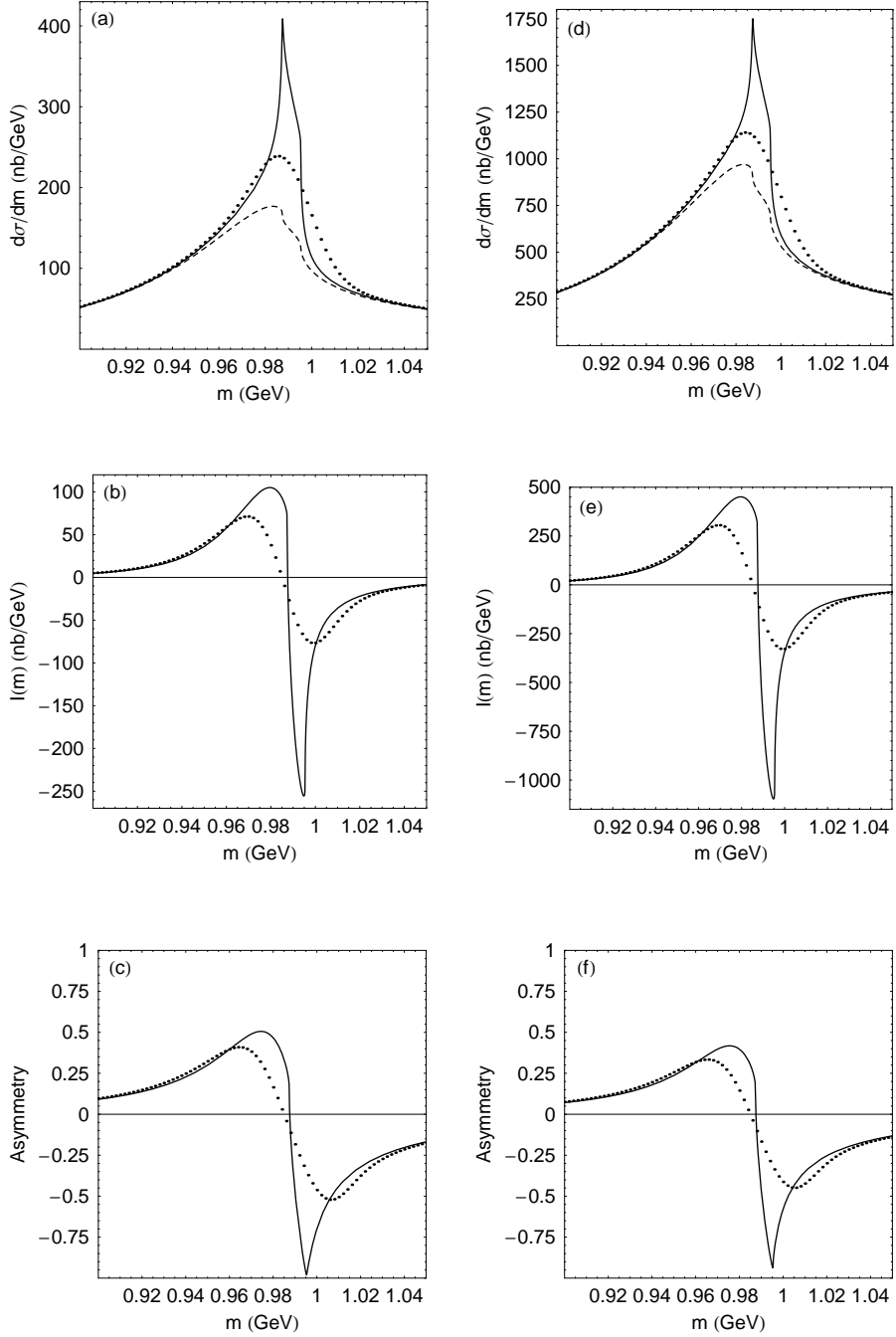}}
\caption{Manifestation of the $a^0_0(980)-f_0(980)$ mixing effect
in the reaction $\pi^-p\to a^0_0 (980)n\to (\eta\pi^0)_S\,n$ on a
polarized target at $P^{\pi^-}_{lab}=18.3$ GeV in the $\rho_2$ and
$\pi$ exchange model. The solid curves in (a),(b),(c) show
$d\sigma/dm$, $\,I(m)$, for the interval $0\leq-t\leq0.025$
GeV$^2$, and the corresponding asymmetry
$A(0\leq-t\leq0.025\mbox{\,GeV}^2,\,m)$, respectively.  The dashed
curve in (a) shows the $\rho_2$ exchange contribution to
$d\sigma/dm$. The dotted curves in (a),(b),(c) show $d\sigma/dm$,
$I(m)$, smoothed with a Gaussian mass distribution with the
dispersion of ten MeV, and the corresponding asymmetry,
respectively. Plots (d),(e),(f) show the same as plots (a),(b),(c)
but for the interval $0\leq-t\leq0.2$ GeV$^2$. The overall sign of
the $I(m)$ and, consequently, the asymmetry is unknown and was
chosen arbitrarily.}\end{figure}

\begin{figure}\centerline{\epsfysize=7.5in\epsfbox{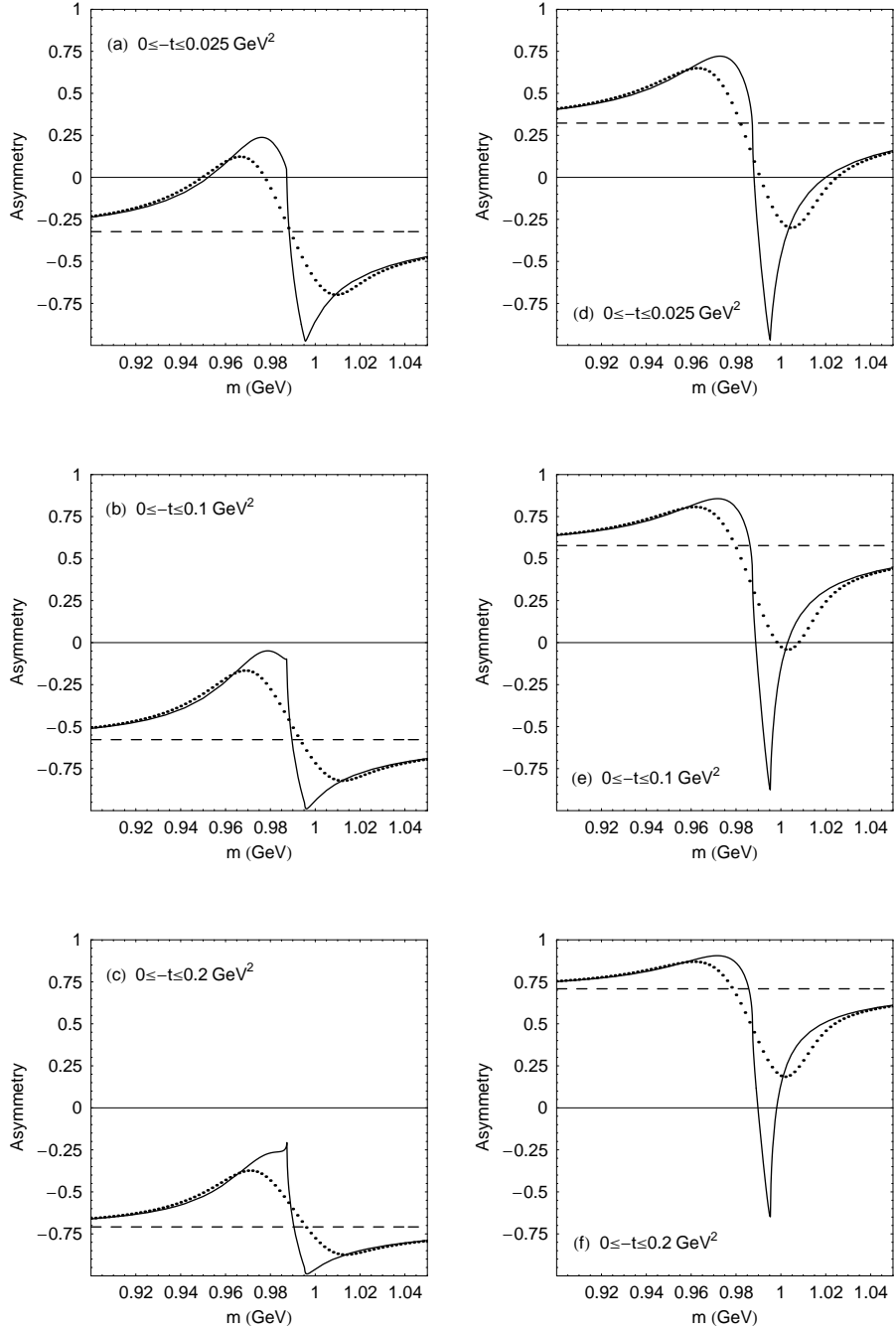}}
\caption{The solid (and dotted ) curves in (a),(b),(c) show the
asymmetry $A(0\leq-t\leq-t_2, \,m)$ for the different intervals of
$-t$ in the $\rho_2$, $\pi$, and $b_1$ exchange model without (and
with) a Gaussian mass smearing (with the dispersion of ten MeV).
The dashed lines correspond to the asymmetry in the $\rho_2$ and
$b_1$ exchange model. The overall sign of the asymmetry and the
relative sign of the $b_1$ and $\pi$ exchange amplitudes are
unknown and were chosen arbitrarily. Plots (d),(e),(f) show the
same as  plots (a),(b),(c) but for the different choice of the
sign of the $b_1$ exchange amplitude.}\end{figure}

\begin{figure}\centerline{\epsfysize=7.5in\epsfbox{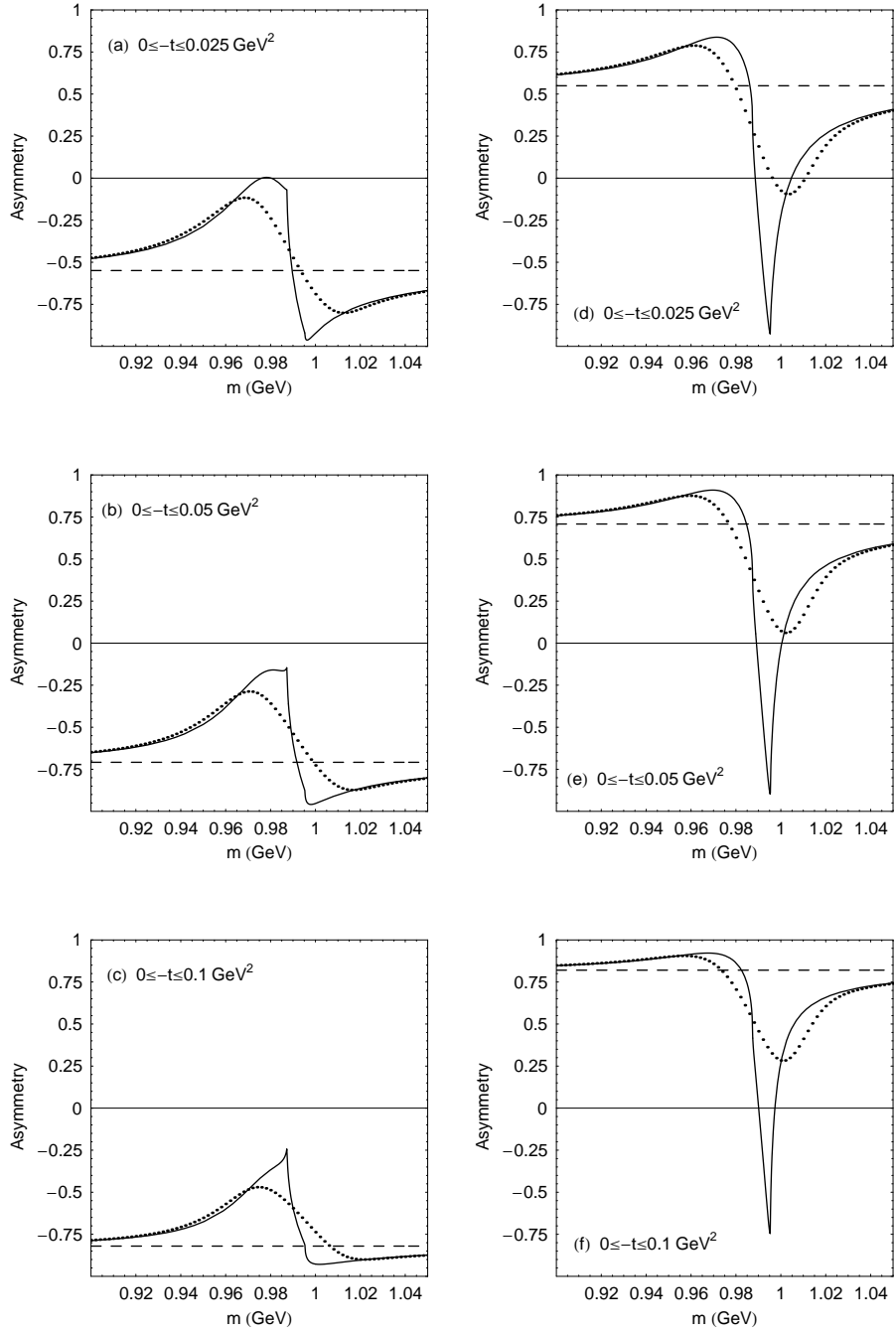}}
\caption{The same as in Fig. 6 but for the $\rho_2$ and $b_1$
exchange contributions shown in Fig. 4.}\end{figure}

\end{document}